\documentclass[%
 reprint,
 superscriptaddress,
 amsmath,amssymb,
 aps,
]{revtex4-2}

\usepackage{graphicx}
\usepackage{dcolumn}
\usepackage{bm}
\usepackage{comment}
\usepackage{bm, color, ulem}
\usepackage{gensymb}
\usepackage{siunitx}
\usepackage{braket}
\usepackage{amsmath}

\usepackage{xcolor}

\begin{document}

\preprint{APS/123-QED}

\title{Long-lived coherences in strongly interacting spin ensembles}

\author{William K. Schenken}
\altaffiliation{These authors contributed equally.}
\affiliation{Department of Physics, University of Colorado, Boulder 80309 CO, USA}%
\affiliation{Department of Physics, University of California, Santa Barbara 93106 CA, USA}%

\author{Simon A. Meynell}
\altaffiliation{These authors contributed equally.}
\affiliation{Department of Physics, University of California, Santa Barbara 93106 CA, USA}%

\author{Francisco Machado}
\affiliation{ITAMP, Harvard-Smithsonian Center for Astrophysics, Cambridge, Massachusetts, 02138, USA}
\affiliation{Department of Physics, Harvard University, Cambridge 02138 MA, USA}%

\author{Bingtian Ye}
\affiliation{Department of Physics, Harvard University, Cambridge 02138 MA, USA}%

\author{Claire A. McLellan}
\affiliation{Department of Physics, University of California, Santa Barbara 93106 CA, USA}%

\author{Maxime Joos}
\affiliation{Department of Physics, University of California, Santa Barbara 93106 CA, USA}%

\author{V. V. Dobrovitski}
\affiliation{QuTech, Delft University of Technology, Lorentzweg 1, 2628 CJ Delft, The Netherlands}%
\affiliation{Kavli Institute of Nanoscience, Delft University of Technology, Lorentzweg 1, 2628 CJ Delft, The Netherlands}

\author{Norman Y. Yao}
\affiliation{Department of Physics, Harvard University, Cambridge 02138 MA, USA}%

\author{Ania C. Bleszynski Jayich}
\affiliation{Department of Physics, University of California, Santa Barbara 93106 CA, USA}%

\date{\today}

\begin{abstract}

Periodic driving has emerged as a powerful tool to control, engineer, and characterize many-body quantum systems. 
However, the required pulse sequences are often complex, long, or require the ability to control the individual degrees of freedom. 
In this work, we study how a simple Carr-Purcell Meiboom-Gill (CPMG)-like pulse sequence can be leveraged to enhance the coherence of a large ensemble of spin qubits and serve as an important characterization tool.
We implement the periodic drive on an ensemble of dense nitrogen-vacancy (NV) centers in diamond and examine the effect of pulse rotation offset as a control parameter on the dynamics. 
We use a single diamond sample prepared with several spots of varying NV density, which, in turn, varies the NV-NV dipolar interaction strength.
Counter-intuitively, we find that rotation offsets deviating from the ideal $\pi$-pulse in the CPMG sequence (often classified as pulse errors) play a critical role in preserving coherence even at nominally zero rotation offset. 
The cause of the coherence preservation is an emergent effective field that scales linearly with the magnitude of the rotation offset.
In addition to extending coherence, we compare the rotation offset dependence of coherence to numerical simulations to measure the disorder and dipolar contributions to the Hamiltonian to quantitatively extract the densities of the constituent spin species within the diamond.

\end{abstract}

\maketitle

Engineered quantum systems have emerged as powerful and flexible tools for probing many-body physics.
Whether composed of atomic, superconducting, or solid-state defect degrees of freedom, these platforms now routinely provide important insights into myriad phenomena such as phases of matter~\cite{Abanin2019_MBL_thermal_engtangle, Smith2016_MBL_QuantumSimulator, Yao2018_dipolar_spin_liquid, Choi2017_TimeCrystal_DisorderedDipolar,Kucsko2018_CriticalThermalization} and quantum decoherence~\cite{Davis2023,Dwyer2022_ProbingSpinDynamics,Decoherence_Bauch_2020,Buchleitner2003_InteractionInducedDecoherenceAtomic,Takahashi2011_molecularmagnets}, and can even enable the generation of metrologically useful entangled states \cite{Zheng2022_MetrologicalStates,Chu2023_QuantumMetrologicalLimit,hines2023_SpinSqueezeRydberg,bornet2023_SpinSqueezeRydberg,eckner2023_SpinSqueezeRydberg,franke2023_SpinSqueezeRydberg}.
The physics of these many-body phenomena crucially depends on the interplay between two distinct types of interactions: those within the system itself (internal) and those between the system and its environment (external).
Hence, controlling and characterizing these interactions is central to developing a many-body quantum simulator or a sensor.

\begin{figure}
    \includegraphics[width =3.3in]{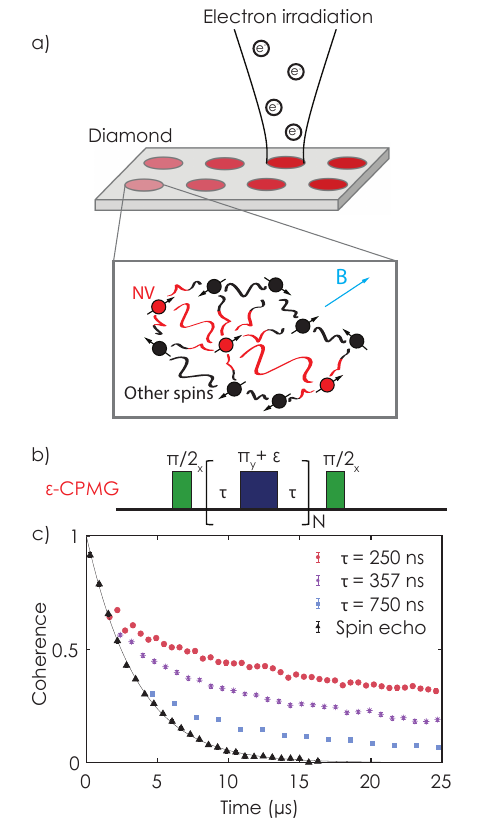}
    \caption{
    Measurement of long-lived coherences in a dense NV ensemble. 
    a) Cartoon sketch of our spin system, a CVD-grown, nitrogen-doped diamond sample showing spots that have undergone localized electron irradiation of varying dosage, resulting in high NV densities that vary across spots. 
    The NV ensemble (red) is mixed randomly with a bath of spins (black). A magnetic field (blue) is applied along the 111 direction. 
    b) The $\epsilon$-CPMG pulse sequence. 
    c) Long-lived coherences observed under $\epsilon$-CPMG. For the data shown, $\theta\approx\pi$, and $N$ is varied for various fixed values of $\tau$. For comparison, the spin echo is also shown ($N$=1, $\theta\approx\pi$, variable $\tau$). 
    Data is taken on spot A (see Table \ref{table:spots} and Fig. \ref{fig:FigHighVsLowDensity_and_VariousN}(a)).  
    Details of the fit to the spin echo data are given in the supplemental information \cite{supplemental_note}. 
    }
    \label{fig:FigMain}
\end{figure}

Defects in semiconductors such as diamond - in particular, the nitrogen-vacancy (NV) center - are especially promising as a platform for exploring many-body physics owing to their optical polarizability and their ability to access large system sizes \cite{Zu2021_emergentHD,Davis2023,Wei2015,Choi2017_TimeCrystal_DisorderedDipolar}. 
While it has recently become possible to synthesize ensembles of dipolar interacting NV centers \cite{Eichorn_Optimizing_2019,Quantum_Zhou_2020,Davis2023}, additional interactions with the environment often destroy the system's coherence, limiting the landscape of explorable quantum many-body phenomena. 
For example, other defects in the diamond lattice, such as the spin-1/2 substitutional nitrogen impurity (P1 centers) or NV centers of other orientations, introduce a local, random, fluctuating magnetic field; such fluctuations are typically the primary source of decoherence in NV systems.
To design and realize novel many-body states in NV ensembles and gain a clear understanding of decoherence pathways, it is important to characterize the various interactions within the system and between the system and its environment. 
Armed with this information, suitable protocols can be used to either mitigate \cite{Joos2022_ProtectingCoherence, Bluvstein2019_ExtendingCoherence} or leverage \cite{Zhang2017_ObservationDTC} these interactions.

\begin{figure}
    \centering
    \includegraphics[width=3.3in]{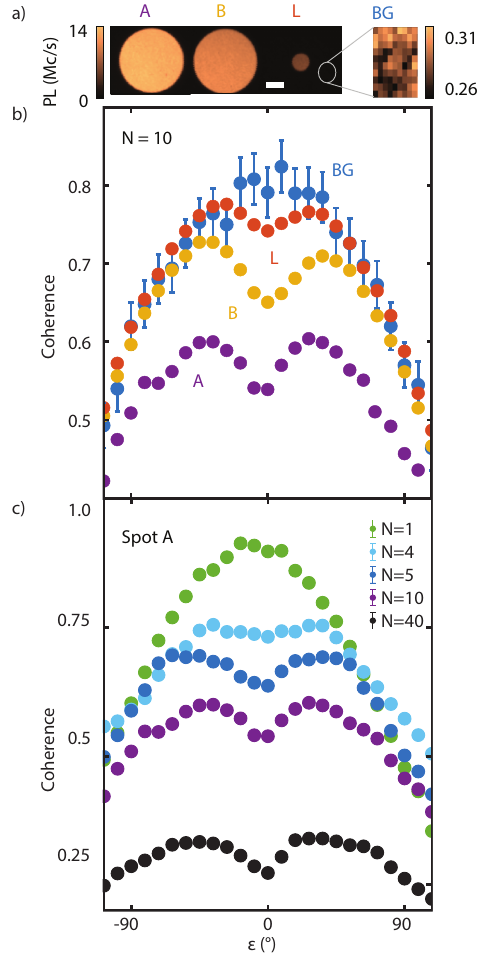}
    \caption{Behavior of $\epsilon$-CPMG sequence. (a) Confocal images of spots A, B, and L, showing a range of brightness indicating a range of NV densities. The scale bar is $\SI{5}{\micro\meter}$. b) Coherence versus $\epsilon$ for various NV densities at fixed $N = 10$ and $\tau = \SI{250}{\nano\second}$, and (c) for various $N$ for a given NV density (spot A) and $\tau$. 
    See the supplemental information \cite{supplemental_note} for a comparison of this data to an analytic model.
    }
    \label{fig:FigHighVsLowDensity_and_VariousN}
\end{figure}

One powerful method to probe and manipulate interactions in spin systems is Hamiltonian engineering, where the spins are manipulated via coherent pulses on a timescale faster than any interaction strength present in the system, resulting in an effective time-averaged Hamiltonian $\bar{H}$ that describes the evolution of the spins.
Through fine-tuning of the pulse sequence parameters (i.e., rotation angle, rotation axes, pulse duration, timing between pulses), the effective Hamiltonian can be engineered 
to extend coherence by decoupling external disorder (e.g., Hahn echo \cite{HahnEcho_1950} or CPMG \cite{CP_1954, MG_1958}) or suppressing internal interactions (e.g., WAHUHA \cite{WAHUHA_1968}). 
Recently, complex sequences have been designed that are capable of suppressing both disorder and interactions while being robust to pulse errors (DROID \cite{Robust_Choi_2020}).
Outside of dynamical decoupling, Hamiltonian engineering has also been applied to simulating many-body physics in engineered spin systems \cite{Zhang2017_ObservationDTC}.

In this Letter, we demonstrate how the continuous tuning of a simple CPMG-like pulse sequence, hereafter referred to as $\epsilon$-CPMG, can be leveraged to preserve coherence out to timescales approaching $T_1$ in the rotating frame, $T_{1,\rho}$, regardless of the ratio of dipolar interactions to disorder. 
Using $\epsilon$-CPMG, we measure the strength of dipolar interactions and disorder, an important characteristic of many-body systems.
By fitting the data to numerical simulations, we quantitatively estimate the density of NV system spins and the surrounding spin bath.
By sweeping rotation offset, we tune the dynamics from being disorder-dominated to dipole-dipole dominated in sufficiently dense samples. 
Importantly, because $\epsilon$-CPMG \textit{leverages} rotation offsets, it is inherently robust to rotation offset pulse errors, a feature that contrasts it with most other pulse sequences that necessitate increased complexity to achieve pulse error robustness.

We characterize a diamond sample with regions of different densities of NVs tuned via electron irradiation (Fig. \ref{fig:FigMain}(a)) whose dynamics are governed by the following Hamiltonian,
\begin{equation}
    H = \sum_i B^z_i(t) \sigma_i^z + \sum_{i<j} J_{ij} \left( \sigma^x_i\sigma^x_j + \sigma^y_i\sigma^y_j - \sigma^z_i\sigma^z_j \right),
    \label{eqn:Hamiltonian}
\end{equation}
where the first summand is referred to as disorder or external interactions, the second summand contains the dipolar or internal interactions, $B^z_i(t)$ is the local onsite field at spin $i$, which can vary as a function of time, $t$. The dipolar coupling between spins $i$ and $j$ is $J_{ij}$, and $\sigma^{x,y,z}_i$ are the Pauli spin operators for the $i$th spin. 
Our engineered Hamiltonian approach accomplishes this using the $\epsilon$-CPMG sequence (schematically shown in Fig.~\ref{fig:FigMain}(b)), which varies a single parameter --- the pulse rotation offset $\epsilon$, defined as the deviation from a perfect $\pi$-pulse.
As we will demonstrate, this simple sequence permits the characterization of the internal and external interactions of the NV samples without necessitating either arbitrary waveform generators for fast and precise phase control or multiple microwave drives as in double-resonance techniques, such as in double electron-electron resonance (DEER).
Fig.~\ref{fig:FigMain}(c) shows an example of these long-lived coherences under the application of the $\epsilon$-CPMG sequence.

We investigate a chemical vapor deposition (CVD)-grown diamond sample (sample C041) with a several-$\SI{}{\micro\meter}$ thick nitrogen-doped layer, as detailed in previous work \cite{Eichorn_Optimizing_2019}. 
As schematically indicated in Fig.~\ref{fig:FigMain} (a), the diamond was irradiated in several $\sim\SI{}{\micro\meter}$-scale spots with electrons of varying dosage and energy. 
The electron irradiation creates vacancies, and an 850$\degree$C anneal promotes the conversion of nitrogen to NV centers. 
A key feature of our sample is that in the spots with higher irradiation dosage, the NV density is high enough to play a significant role in the decoherence dynamics \cite{Eichorn_Optimizing_2019}. 
Here we examine three distinct irradiated spots on sample C041, whose details are shown in Table~\ref{table:spots}. 
We also study the unirradiated background, where the NV density is sufficiently low such that NV-NV interactions are negligible.
All measurements presented are performed in a $\sim320$~G magnetic field aligned with one of the four NV axes. Microwave (MW) pulses address the $\{|0\rangle, \left |-1\right\rangle\}$ transition of the aligned NV group. 

\begin{table*}
\begin{center}
\begin{tabular}{|c | c | c | c | c | c | c | } 
 \hline
 Spot & Dosage & Energy & [NV]$_\textrm{ID}$ & [P1]$_{\textrm{DEER}}$ & [NV]$_{\epsilon-\textrm{CPMG}}$ & [spin-defect]$_{\epsilon-\textrm{CPMG}}$   \\ [0.5ex] 
 \hline
 \hline
 A & $10^{21}$ (e$^-$/cm$^2$) & 200 keV & $2.7\pm0.08 ~\mathrm{ppm}$ & $3.8\pm0.2 ~\mathrm{ppm}$ & $2.1\pm0.3~\mathrm{ppm}$ & $23.2\pm0.5~\mathrm{ppm}$\\ 
 B & $10^{20}$ (e$^-$/cm$^2$) & 200 keV & $2.2\pm0.21~\mathrm{ppm}$ & $10.5\pm0.2 ~\mathrm{ppm}$ & $1.3\pm0.4~\mathrm{ppm}$ & $17.4\pm0.9~\mathrm{ppm}$ \\
 L & $10^{22}$ (e$^-$/cm$^2$) & 145 keV & $1.0\pm0.2~\mathrm{ppm}$ & - & $0.6\pm0.3~\mathrm{ppm}$ & $16.1\pm0.7~\mathrm{ppm}$ \\
\hline
\end{tabular}
\caption{Comparison of spin densities for spots A, B, and L as obtained from instantaneous diffusion (ID), DEER (Ref. \cite{Eichorn_Optimizing_2019}) and from $\epsilon$-CPMG, showing good agreement in the NV density. Explanations for the discrepancies are discussed in the text.}
\label{table:spots}
\end{center}
\end{table*}

The pulse sequence is shown schematically in Fig.~\ref{fig:FigMain}(b).
After optical polarization into $\ket{0}$ using a $\SI{532}{\nano\meter}$ green laser, a microwave $\pi/2$ pulse in the $x$-direction initializes the spins into the state $\ket{+y}$. 
Then, with a CPMG-like train of pulses, the spins undergo repeated rotations about the $+\hat{y}$ axis by an angle $\theta = \pi + \epsilon$. 
At the end of the sequence, the NVs are mapped into a population by a final $\pi/2$ pulse along the $x$-direction, and the state is read out optically via spin-dependent photoluminescence under a $\SI{532}{\nano\meter}$ laser pulse.
The resulting effective Hamiltonian under the train of pulses has particularly simple intuition for two specific extremal values of $\epsilon$: when $\epsilon=0$, i.e., a conventional CPMG sequence is applied, static disorder is decoupled~\cite{HahnEcho_1950, CP_1954, MG_1958, Zhang2007_ModellingDecoherence} and dipolar couplings are unaffected (the dipolar interaction between two spins is invariant under a $\pi$ rotation of both spins); when $\epsilon = \pm \pi/2$, the dipolar interactions between NVs are maximally averaged out~\cite{Multiple_Ostroff_1966} while the static disorder is only averaged out half as effectively as in the $\epsilon=0$ case.
When both static disorder and dipolar interactions are present, an intermediate $\epsilon$ is best suited to maximizing the coherence in such cases. 
Measuring NV ensemble coherence as a function of $\epsilon$ determines the strength of dipolar coupling relative to disorder.

In Figure~\ref{fig:FigHighVsLowDensity_and_VariousN}, we plot the coherence of multiple different NV ensembles subject to our $\epsilon$-CPMG pulse sequence as a function of $\epsilon$ for $\tau = \SI{250}{\nano\second}$. 
The results demonstrate how this one simple knob can be tuned to optimize coherence for different spin environments, realized in the differently irradiated spots (Fig.~\ref{fig:FigHighVsLowDensity_and_VariousN}(a) and Table~\ref{table:spots}).
Fig.~\ref{fig:FigHighVsLowDensity_and_VariousN}(b) shows that in the lowest NV density spot (spot BG) coherence is maximized for $\epsilon=0\degree$, as in an ideal CPMG sequence. 
At $\epsilon=90\degree$ the coherence drops to about half its maximum value. 
As the NV density increases in spots L, B, and A, internal dipolar interactions start to dominate the decoherence dynamics at small $\epsilon$, leading to a double-humped feature where coherence is maximized at a nonzero $\epsilon$. 

Having understood the features of the sequence at large $N$, we next examine the system's coherence as a function of the number of pulses, $N$.
Fig.~\ref{fig:FigHighVsLowDensity_and_VariousN}(c) shows coherence vs. $\epsilon$ in spot A for a varying number of pulses, $N$, while keeping the interpulse spacing constant. 
Qualitatively, we observe the emergence and deepening of a coherence dip at $\epsilon=0\degree$ as $N$ increases. 
Coherence is also lost at higher $N$ due to the increased duration of the pulse sequence.
In our case, the dip is clearly visible after $\sim5$ pulses.
Using a simple analytical model to compute the system's late-time coherence \cite{supplemental_note}, we reproduce the experimental observations and find good agreement with the relative strength between interactions and disorder.
We find that the periodic drive gives rise to an $\epsilon$-dependent effective magnetic field along the $y$-direction that suppresses depolarization.
In dense ensembles of nuclear spins, Refs. \cite{ajoy2020dynamical, Ridge2014_LongLivedEchos} have observed behavior similar to that shown in Fig.~\ref{fig:FigHighVsLowDensity_and_VariousN}.

\begin{figure}
    \centering
    \includegraphics[width=3.3in]{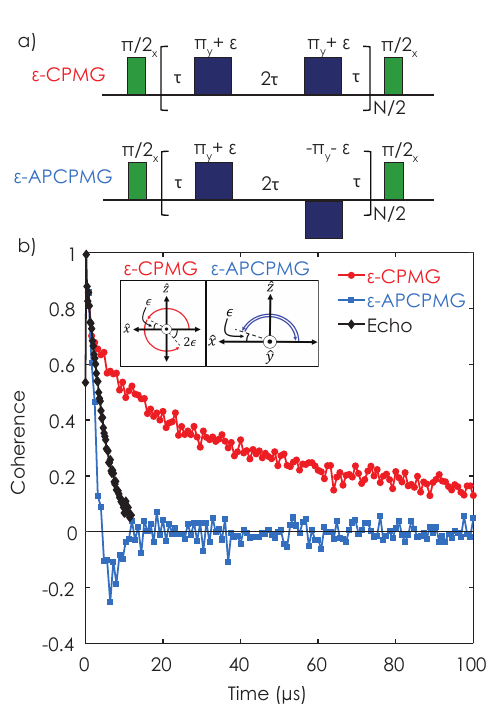}
    \caption{
    Comparison of CPMG to APCPMG, demonstrating the importance of a finite $\epsilon$ for preserving coherence in the $\epsilon$-CPMG sequence.
    a) Schematic depiction of the pulse sequences, CPMG and APCPMG. 
    In both cases, there exists a finite rotation offset but APCPMG is designed to cancel the effect of accumulating rotation offsets.
    b) Experimental data showing the difference in coherence times of the CPMG (red) and APCPMG (blue) sequences with a target of $\epsilon = 0$ using a fixed interpulse spacing of $\tau = \SI{250}{\nano\second}$.  
    A Hahn echo sequence is shown in black (variable interpulse spacing, $N = 1$). 
    The decay of the APCPMG sequence is similar to the Hahn echo, though it has an additional modulation. 
    The inset shows the different dynamics of the NV spin on the Bloch sphere under the two sequences.
    }
    \label{fig:CompareSequence}
\end{figure}

The important role $\epsilon$ plays in the long-lived coherence is further elucidated in Fig.~\ref{fig:CompareSequence}. 
Here we compare the results from the CPMG sequence to an alternating-phase CPMG (APCPMG) sequence applied to spot A (Fig.~\ref{fig:CompareSequence}a). These two sequences have the same filter function \cite{Biercuk2011_DynamicalDecouplingFilterDesign,Degen2017_Sensing,Cywinski2008_DephasingTime} but differ in that APCPMG is explicitly designed to cancel the effect of accumulating rotation offsets to lowest order in $\epsilon$.
Perhaps counterintuitively, the CPMG sequence (red) shows substantially longer coherence times than the APCPMG sequence (blue), further evidencing the important role of finite rotation offset in the extension of coherence.
Plotted in black is the Hahn echo coherence, which decays on a timescale similar to that of the APCPMG sequence, indicating that the disorder is largely quasi-static; a Hahn echo sequence, with just a single $\pi$ pulse, is sufficient to decouple disorder.
We thus conclude that cumulative rotation offsets are, in fact, key to the coherence extensions observed here. 
In nuclear spin systems, finite pulse duration has been used to explain a similar difference between CPMG and APCPMG \cite{Unexpected_Li_2007, DongControllingCoherence2008}.
In the following analysis, we will include the effects of both finite pulse duration and rotation offset.

While an approximate effective Hamiltonian description of the coherence data, as detailed in the supplemental information \cite{supplemental_note}, provides physical intuition, it fails to provide a direct and quantitative mapping between the properties of the sample and the observed dynamics.
We tackle this gap by numerically simulating the full dynamics of the NV ensemble interacting with a bath of spin-1/2 defects.
Crucially, this approach enables us to capture the full extent of the experimental observations (both as a function of angle error $\epsilon$ and number of pulses $N$) and use the agreement between numerical data and experiments to quantitatively characterize important features of the sample such as the density of different spin defects, thus revealing the relative strengths of disorder and interactions.

To this end, we numerically compute the dynamics of an NV ensemble with density $n_{\mathrm{NV}}$ surrounded by a bath of defects at density $n_{\mathrm{bath}}$.
The NVs are all initialized in the $\ket{+y}$ orientation, and their subsequent dynamics have three contributions: 1) dipolar interactions between NVs, 2) Larmor precession arising from local, random magnetic fields generated by the bath spins, and 3) periodic rotations by $\pi+\epsilon$ about the $\hat{y}$ axis.  
By contrast, the bath spin defects are assumed to be randomly polarized in either $\pm 1/2$ and exhibit no coherent dynamics.
Instead, we consider their dynamics as a stochastic process that flips the polarization of the spin defect with some characteristic time scale, giving each spin in the bath a correlation time $\tau_c$ that is related to the density of the spin bath, as described in Ref. \cite{Decoherence_Bauch_2020}.
Such bath dynamics are crucial to capture the observed decoherence dynamics of the NV centers - without them, the effect of the spin bath can be exactly canceled for perfect $\pi$ pulses.
In addition, we incorporate different experimental effects into the numerics, such as finite pulse duration, which are difficult to incorporate in a simple theoretical analysis \footnote{Average Hamiltonian theory has been used to show that a finite pulse duration gives rise to effects similar to those studied here. For a detailed analysis of the effect of finite pulses, see Ref. \cite{Unexpected_Li_2007}}.

By computing the NV ensemble dynamics over different sets of densities of NVs and bath spins $\{n_{\mathrm{NV}}, n_{\mathrm{bath}}\}$ and averaging over different spatial configurations of the NV and bath ensembles \cite{Hahn_LongLived_2021, Feldman_Configurational_1996}, we obtain the coherence dynamics for a wide range of samples as a function of $\epsilon$ and $N$. 
We compare the resulting coherence dynamics across different $N:N>3$\footnote{The observations for the first few cycles,  $N\leq 3$, are expected to depend on the details of the initialization process, which we do not capture in our simulation.} for each pair of parameters, $\{n_{\mathrm{NV}}, n_{\mathrm{bath}}\}$, estimating the agreement between numerical and experimental data via a $\chi^2$-like measure.

\begin{figure}
    \centering
    \includegraphics[ width=3.3in]{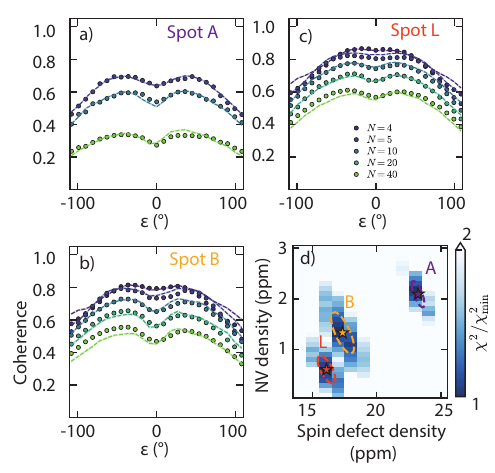}
    \caption{Simulations of long-lived coherences. (a), (b), (c): Coherence versus $\epsilon$ for various $N$ and for spots A, B, L. Markers show experimental data. Plotted as dashed lines are the numerics for the parameters $\{\rho_{NV}, \rho_{\mathrm{bath}}\}$ that minimized $\chi^2$. (d) $\chi^2$ contour plots as a function of $\rho_{NV}$ and $\rho_{\mathrm{bath}}$. The colormaps for each spot have been superimposed. Separated colormaps are shown in the supplemental\cite{supplemental_note}.}
    \label{fig:Numerics}
\end{figure}

Sample fits and $\chi^2$ contour plots are shown in Fig.~\ref{fig:Numerics}.  
We find that there is a clear region within this parameter space that minimizes $\chi^2$. 
The corresponding densities are shown in Table \ref{table:spots}. The extracted NV densities match the trends of those measured on the same sample via DEER and instantaneous diffusion \cite{Eichorn_Optimizing_2019}, with the precise values falling within or just outside their respective error bars.  
The spin bath density reported here is larger than the P1 density obtained via DEER, primarily because the present method is sensitive to all spinful defects in the diamond, not just to P1s, whereas DEER is, by design, only sensitive to a single group of P1 centers. 
These extra spinful defects could be a result of the high irradiation dosage. 
Our simulations also give an NV density slightly lower than what was reported previously from instantaneous diffusion measurements \cite{Eichorn_Optimizing_2019}. 
The previous analyses necessitated several approximations, such as a quasi-static spin bath in the case of DEER and a dominant NV bath in the case of instantaneous diffusion. 
The present numerical treatment is free from these assumptions and hence should be a more reliable probe of the NV density.

We stress that our ability to quantitatively extract the densities of competing disordered and dipolar spin baths harbors key advantages over other methods such as DEER, XY8, and instantaneous diffusion, namely technical simplicity, robustness to pulse errors in the form of rotation offsets, and the ability to simultaneously probe both NV density and disorder without assuming one to be dominant.

DEER decay is a common technique requiring two microwave sources that can measure specific species' densities as it offers spectral resolution; however, obtaining a figure of merit for the total disordered spin bath can be challenging exactly because DEER requires addressing the individual baths spectrally and so spectrally broad baths may be difficult to address.
In addition to this, while NV-NV DEER is possible, it is challenging because of the relatively low density of NV centers in a single group; attempts to increase the density of probed NV spins by probing three groups simultaneously are thwarted by the different alignment of the rf field to each of the three groups, and it is required to wait a time longer than $T_2^{*}$ for the bath spins to dephase.
Instantaneous diffusion can be used to measure the strength of the dipole-dipole interaction via the extension of $T_2$ under non-$\pi$ pulses but cannot easily measure disorder and assumes that the change in $T_2$ with pulse angle is unrelated to the change in coupling to disorder.
Using decoupling techniques like XY8 or DROID \cite{Quantum_Zhou_2020} and comparing the coherence times between dipole-decoupled and disorder-decoupled is another way to assess the relative importance between these terms. 
In contrast to the sequence presented in this Letter, XY8, and DROID require complicated sequencing; in addition, extracting quantitative values for the strength of disorder vs dipolar interactions is complicated.

In this Letter, we have shown how the $\epsilon$-CPMG sequence extends the coherence of spin ensembles to times approaching $T_{1,\rho}$ and provides quantitative information about the relative strengths of dipolar interactions and disorder, important figures of merit for many-body systems. 
Further, we show how our approach to Hamiltonian engineering can tune from dipolar-dominated to disorder-dominated physics.
Using this tunability, we identify an optimal $\epsilon$ for maximizing coherence of an NV ensemble.
We benchmark our pulse sequence, its ability to tune interaction strengths, and our numerical methods for estimating spin densities on a single sample with multiple different NV densities.

As decoherence mitigation becomes an increasingly large scientific endeavor, fast, straightforward, and accurate diagnostic sequences like the one presented here will become ever more essential for understanding how many-body quantum systems decohere.
Fast and efficient characterization of quantum systems may become increasingly important in the `pipeline' for creating and developing samples; we propose that the $\epsilon$-CPMG sequence represents one such way of mitigating this potential bottleneck because it is efficient, has a small technical footprint, and is unambiguous in its diagnosis.
Moreover, while we expect that $\epsilon$-CPMG will be primarily applicable to solid-state systems, we note that the treatment is generic and could be applied to systems beyond dense NV ensembles - in any disordered system where the nature of interactions between particles within the system and without must be understood, we expect $\epsilon$-CPMG to be useful.
Finally, we anticipate that this framework may be extended beyond a two-component Hamiltonian by introducing additional tuning parameters within a periodic drive (for example, $\epsilon_x$ for $x$-pulses and $\epsilon_y$ for $y$-pulses) to extend this treatment to even more complicated Hamiltonians.

We gratefully acknowledge support of the Army Research Office through the MURI program grant number W911NF-20-1-0136, the U.S. Department of Energy BES grant No. DE-SC0019241, and the DARPA DRINQS program (contract D18AC00015KK1934 and D18AC00033).
This work is part of the research programme NWO QuTech Physics Funding (QTECH, programme 172) with project number 16QTECH02, which is (partly) financed by the Dutch Research Council (NWO). W.K.S. acknowledges support from the NDSEG Fellowship.
S.A.M. acknowledges support from the UCSB Quantum Foundry (NSF DMR-1906325) and support from the Canada NSERC (Grant No. AID 516704-2018).
F.M. acknowledges support from the NSF through a grant for ITAMP at Harvard University.

\bibliography{apssamp}

\end{document}


\preprint{APS/123-QED}

\title{Long-lived coherences in nitrogen-vacancy centers in interacting spin ensembles: Supplemental information}

\author{William K. Schenken}
\altaffiliation{These authors contributed equally.}
\affiliation{Department of Physics, University of Colorado, Boulder 80309 CO, USA}%
\affiliation{Department of Physics, University of California, Santa Barbara 93106 CA, USA}%

\author{Simon A. Meynell}
\altaffiliation{These authors contributed equally.}
\affiliation{Department of Physics, University of California, Santa Barbara 93106 CA, USA}%

\author{Francisco Machado}
\affiliation{ITAMP, Harvard-Smithsonian Center for Astrophysics, Cambridge, Massachusetts, 02138, USA}
\affiliation{Department of Physics, Harvard University, Cambridge 02138 MA, USA}%

\author{Bingtian Ye}
\affiliation{Department of Physics, Harvard University, Cambridge 02138 MA, USA}%

\author{Claire A. McLellan}
\affiliation{Department of Physics, University of California, Santa Barbara 93106 CA, USA}%

\author{Maxime Joos}
\affiliation{Department of Physics, University of California, Santa Barbara 93106 CA, USA}%

\author{V V Dobrovitski}
\affiliation{QuTech, Delft University of Technology, Lorentzweg 1, 2628 CJ Delft, The Netherlands}%
\affiliation{Kavli Institute of Nanoscience, Delft University of Technology, Lorentzweg 1, 2628 CJ Delft, The Netherlands}

\author{Norman Y. Yao}
\affiliation{Department of Physics, Harvard University, Cambridge 02138 MA, USA}%

\author{Ania C. Bleszynski Jayich}
\affiliation{Department of Physics, University of California, Santa Barbara 93106 CA, USA}%
\date{\today}

\maketitle

\section{Epsilon-dependence of the coherence}

To gain some intuition for the qualitative shape of the coherence vs. $\epsilon$, we present a theoretical analysis of the coherence after ten pulses.
We analyze each of the three spots using this treatment and show that the strength of the dip corresponds to a term proportional to the variance of the dipole-dipole coupling, $\langle J_{i,j}^2 \rangle$.

To derive an effective, time-independent Hamiltonian, we examine the time-evolution operator over two cycles, given a native Hamiltonian, $H = \sum_i B^z_i \sigma_i^z + \sum_{i<j} J_{ij} \left( \sigma^x_i\sigma^x_j + \sigma^y_i\sigma^y_j - \sigma^z_i\sigma^z_j \right) = H_{\mathrm{on-site}} + H_{\mathrm{dipolar}}$.

\begin{align}
    \hat{U}^2 = e^{-iH\tau}R_y(\pi + \epsilon)e^{-2iH\tau}R_y(\pi + \epsilon)e^{-iH\tau}\\
    =  e^{-iH\tau}R_y(\epsilon)e^{-2i\tilde{H}\tau}R_y(\epsilon)e^{-iH\tau},
\end{align}

where $R_y(\theta)$ is the rotation operator about the $y$-direction by an angle, $\theta$ and $\tilde{H} =R_y(\pi)HR_y(\pi)= -H_{\mathrm{on-site}} + H_{\mathrm{dipolar}} $.
$B_i^z$ is the $z$-field for spin $i$ and assumed to be time-independent for this analysis, for the numerical treatment in the main text, we relax this assumption and allow a time-varying field.
We then find an effective time-independent Hamiltonian using the Magnus expansion focusing on the region close to $\epsilon = 0$ because this is the regime where the effects of finite $J_{i,j}$ are most pronounced.

Our zeroth order term will be given by,

\begin{equation}
    H_{\mathrm{eff}}^{(0)} =  \frac{2\epsilon\sum \sigma^y_i}{4\tau} + H_{\mathrm{dipole}} = B_{\mathrm{eff}}\sum_i^M \sigma^y_i + H_{\mathrm{dipole}}.
\end{equation}

Higher order terms will be generated by the commutators between $H$, $\frac{\epsilon\sigma^y}{\tau}$, and $\tilde{H}$.

We measure the average coherence along the $y$-direction, averaged over the $M$ addressed NV centers. 
After equilibrating, the coherence, $C$, will be given by,

\begin{equation}
\begin{split}
    C = \frac{1}{M}\langle \sum^M_i\sigma^y_i \rangle = \frac{1}{M}\mathrm{tr}\left[ \sum_i^M\sigma^y_i e^{-\beta H_{\mathrm{eff}}}\right]/Z \\
    \approx  \frac{-\beta}{MZ}  \mathrm{tr}\left[\sum^M_i \sigma^y_i H_{\mathrm{eff}}\right] = \frac{-\beta B_{\mathrm{eff}}}{Z}.
\end{split}
\label{eqn:coherence}
\end{equation}

We use, $\rho = e^{-\beta H_{\mathrm{eff}}}/Z$, where $Z$ is the partition function and $\beta$ is given by an effective spin-temperature, $\beta = 1/k_BT$ that satisfies $\beta H_{\mathrm{eff}} \ll 1$.

\begin{figure}
    \centering
    \includegraphics[width=3.3in]{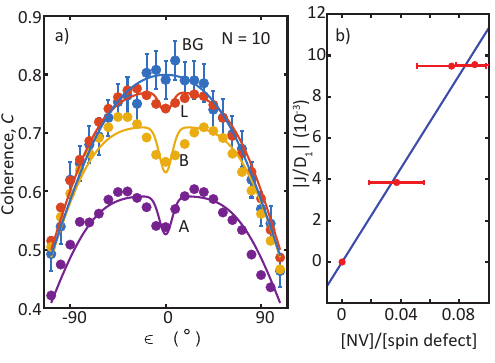}
    \caption{
    (a) Comparison of the coherence as a function of $\epsilon$ across different spots to an analytical model that parameterizes the curves in terms of $J$. 
    (b) A comparison of the fit parameter $J/D_1$ to the ratio of the NV density to the spin defect density obtained via numerical modeling in the main text.
    These have the expected linear relationship.
    }
    \label{fig:SIfits}
\end{figure}

The temperature is determined by the initial value of the $y$ magnetization, which is initialized such that $C = 1$, as detailed in the main text.
We relate the initial and final energies through,

\begin{equation}
\begin{split}
    \sum_i^M B_{\mathrm{eff}} = \langle H_{\mathrm{eff}} \rangle \approx  \frac{-\beta}{Z}  \mathrm{tr}\left[H_{\mathrm{eff}}^2\right].
\end{split}
\label{eqn:initial_final_energy}
\end{equation}

Using Eqn. \ref{eqn:coherence} and \ref{eqn:initial_final_energy}, we arrive at the expression, 

\begin{equation}
\begin{split}
C = \frac{M B^2_\mathrm{eff}}{\mathrm{tr}\left[H_{\mathrm{eff}}^2\right]}.
\end{split}
\label{eqn:simpleCoherence}
\end{equation}

Finally, to arrive at an expression for $\mathrm{tr}\left[H_{\mathrm{eff}}^2\right]$, we examine lowest order terms and expand about $\epsilon = 0$. This results in the following expression,

\begin{equation}
\begin{split}
\mathrm{tr}\left[ H^2\right] = \mathrm{tr}\left[H_{\mathrm{dipolar}}^2 \right] + \mathrm{tr}\left[\sum_i^M (B_\mathrm{eff} \sigma^y_i)^2\right] + \\
\text{higher order terms}\\
 = M(J^2 + B^2_\mathrm{eff}) + \text{higher order terms}\\
  = M(J^2 + D_1^2\epsilon^2 + D_2^2\epsilon^4 + D_3^2\epsilon^6),
\end{split}
\label{eqn:trH2}
\end{equation}

 where $D_{1,2,3}$ are fit coefficients. Using Eqn. \ref{eqn:simpleCoherence} and \ref{eqn:trH2} and simplifying in terms of unique fit parameters, we have:

\begin{equation}
C = \frac{A\epsilon^2}{(J/D_1)^2 + \epsilon^2 + (D_2/D_1)^2\epsilon^4 + (D_3/D_1)^2\epsilon^6 }.
\end{equation}

The typical value of the dipolar coupling $J$, as defined in Eqn. \ref{eqn:trH2}, should be understood as a local value, characterizing only specific region of the sample. 
Indeed, if the quantity  ${\rm tr}{H^2_{\rm dipolar}}$ diverges if averaging is performed over the whole macroscopic 3D ensemble of rare spins \cite{Klauder1962_SpectralDiffusion,Dobrovitski2008_DecoherenceDynamicsSingleVsEnsemble, Hahn_LongLived_2021,Feldman_Configurational_1996}. 
However, this quantity becomes finite and well determined if defined for a finite small region of the sample.

The fits using this model across varying spot densities are shown in Fig. \ref{fig:SIfits} (a).
We use a Gaussian smoothing of $6 \degree$ to account for the random uncertainties in pulse rotation caused by pulse-to-pulse variation and inhomogeneities over the confocal spot.
The parameter $(J/D_1)$ relates the strength of the dipolar contribution to other terms, including $B_{\mathrm{eff}}$ and qualitatively controls the strength of the dip about $\epsilon = 0$. 
We compare $J/D_1$ to the numerically extracted ratio of NV density to spin defect density and find that the two have the expected linear relationship.

\section{Explanation of spin echo fit}

The data in Fig 1(b) was fit to an exponential, $\exp (t/T_2)^n$ with n = 1 and $T_2 = \SI{3.5}{\micro\second}$.
This is consistent with the exponent for Ref. \cite{Davis2023} for high conversion efficiencies where the physics is expected to follow an NV-NV Ramsey type exponential.

\begin{figure}
    \centering
    \includegraphics[ width=3.3in]{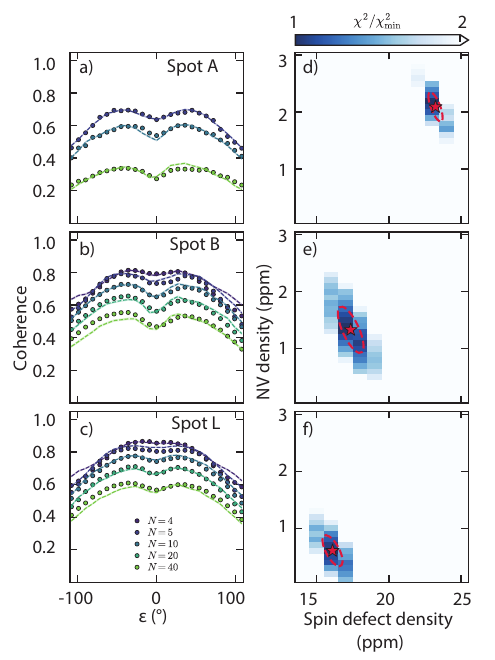}
    \caption{Simulations of long-lived coherences as in the main text with the colormaps, (d), (e), (f) separated instead of superimposed.}
    \label{fig:Numerics}
\end{figure}

\bibliography{apssamp}